\documentclass[cameraready]{Interspeech}
\usepackage{float}
\usepackage{xcolor,colortbl} 
\usepackage{amsmath,amsfonts} %
\usepackage{graphicx}%
\usepackage{textcomp}%
\usepackage{xcolor}%

\usepackage{array}%
\usepackage{pifont}%

\usepackage{amsthm}%
\usepackage{bm}

\usepackage[linesnumbered,ruled,vlined]{algorithm2e}%
\usepackage{setspace}%

\usepackage{multirow}%
\usepackage{siunitx}
\usepackage{footnote}%
\usepackage[utf8]{inputenc}
\usepackage[english]{babel}

\usepackage{blindtext}

\usepackage{dblfloatfix}%

\usepackage{xcolor,colortbl} 
\usepackage{ctable} 
\graphicspath{ {image/} }

\usepackage{pifont}
\newcommand{\xmark}{\ding{55}}

\usepackage{booktabs}
\usepackage{makecell}
\usepackage{comment}
\usepackage[normalem]{ulem}
\usepackage{soul}

\title{\texorpdfstring{CAPS: A \ul{Ca}scaded Reconstruction Model to \ul{P}ower \ul{S}aving in Hearables Using Sub-Nyquist Sampling with Bandwidth Extension}{CAPS: A Cascaded Reconstruction Model to Power Saving in Hearables Using Sub-Nyquist Sampling with Bandwidth Extension}}

\author{Tarikul Islam}{Tamiti}
\author{Sajid Fardin}{Dipto}
\author{Luke}{Baja-Ricketts}
\author{David}{Vergano}
\author{Anomadarshi}{Barua}

\address{
     Cyber-Security Engineering, George Mason University, USA
}

\email{}


\usepackage{comment}

\begin{document}

\maketitle

\begin{abstract}

Hearables are wearable computers worn on the ear. Bone conduction microphones are used with air conduction microphones in hearables for multimodal speech enhancement in noisy conditions.  Despite this potential, current models largely fail to explore how jointly reducing sampling bit resolution and sampling frequency in analog-to-digital converters (ADCs) of hearables impacts both power usage and audio quality. Furthermore, current frameworks cannot do sub-Nyquist sampling in hearables because they lack a method to reconstruct wideband signals from narrowband components. We therefore propose CAPS, which (i) intentionally employs sub-Nyquist sampling and low bit resolution in ADCs, achieving a 3.3x reduction in power consumption in hearables, and (ii) supports streaming operation on mobile platforms with an inference time of 1.36ms and a memory footprint of 11.04MB. CAPS ensures robust speech intelligibility in real-world settings, bridging the gap between efficiency and power savings.

\end{abstract}

\vspace{-1em}
\section{Introduction}
\vspace{-0.4em}

\begin{figure*}[h!]
\vspace{-0.63em}
    \centering
    \includegraphics[width=1.0\textwidth,height=0.15\textheight]{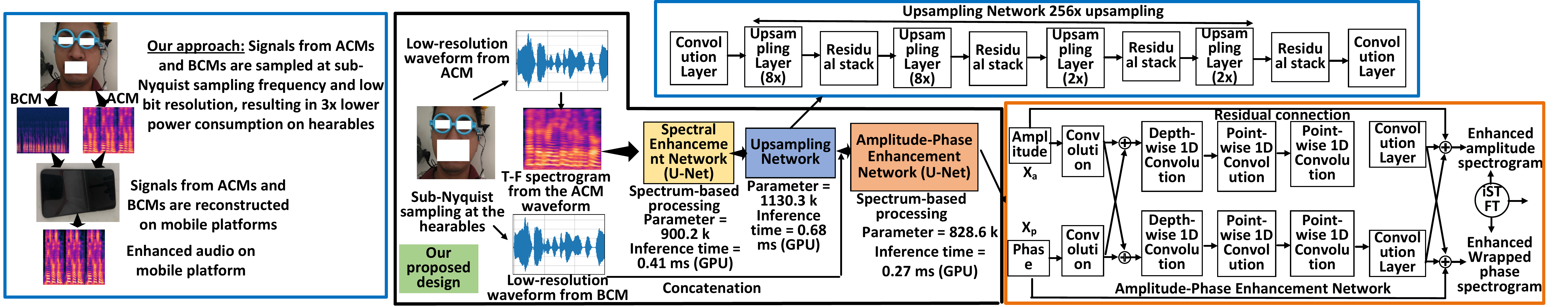}
    \vspace{-01.83em}
    \caption{(Left) Overview of our appraoch. (Middle \& Right) Our proposed design and different modules -- UN \& APEN -- of CAPS.}
    \vspace{-01.99em}
    \label{fig:basic_model}
\end{figure*}

A hearable is a wearable computer that is worn on the ear. Traditionally, air conduction microphones (ACMs) are used in hearables that are prone to background noise. To solve this problem, \textit{bone conduction microphones (BCMs) are commonly used with ACMs as a conditional signal enhancer for multimodal speech enhancement (SE) in noisy conditions \cite{sui2024tramba, tagliasacchi2020seanet, wang2022end}.} However, no SE models  considers the following practical aspects of low-power and low-memory applications in hearables:



\textbf{1. Lowering sampling frequency and bit resolution:} Audio or vibration signals from ACMs and BCMs, respectively, are first sampled at Nyquist rates (greater than 16 kHz) and over 12-bit resolutions by the analog-to-digital converter (ADC). After sampling, audio codecs compress data to reduce the bitrate, saving transmission energy and bandwidth. Later, multimodal SE algorithms are applied on the decompressed data on connected mobile platforms (i.e., cell phones, see Fig. \ref{fig:basic_model} (Left)). \textit{\ul{However, they do not explore how lowering the sampling frequency and bit resolutions in ADCs of hearables jointly impact low-power processing and multimodal SE.}}


\textbf{2. Lacking in multimodal SE methods:} State-of-the-art (SOTA) multimodal SE methods \cite{ma2023clearspeech, han2024earspeech, zhang2025wearse, he2023vibvoice, sui2024tramba,li2022enabling, tagliasacchi2020seanet,mandel2023aero,hauret2023eben,liu2022neural,ho2020denoising,lee2021nu,han2022nu,birnbaum2019temporal,rakotonirina2021self,ma2023clearspeech,he2023vibvoice,kim2023hifi++} have either one or multiple of the following limitations for which they are not suitable for our low-power hearables: \textbf{(i)} The multimodal SE algorithms do not consider \textit{\ul{lower bit resolution and low power applications}} in their frameworks. And, \textbf{(ii)} The joint bandwidth extension (BWE) and SE models are typically evaluated \ul{for single modality (i.e., audio only) and have not been extended to multimodality} (see summary of limitations in rcent works in Table \ref{table:summaryjointBWE}). 

In this paper, we propose CAPS, which \textit{jointly reduces the sampling frequency to \ul{the sub-Nyquist range with lower bit resolutions to reduce power consumption} in hearables and later reconstructs the high-resolution audio in mobile platforms} (see Fig. \ref{fig:basic_model} (Left)). CAPS jointly leverages BWE with multimodal SE \textit{by reducing the sampling frequency and bit resolution from \{24 kHz, 12-bit\} to \{4 kHz, 8-bit\} and achieves a 3.31x reduction in power consumption, ideally increasing the battery life by $\sim$3.31x} for hearables. CAPS is evaluated on both speech and music datasets against six baselines and achieves better performance under noisy conditions with the lowest inference time of 55.11 ms on a mobile platform (i.e., Google Pixel7), which is smaller than the 150 ms threshold set by the International Telecommunication Union \cite{ITU-G114}, indicating its capability of streaming from hearables to mobile platforms. The above results are further verified by using the Galaxy S21.

\begin{table}[ht!]
\scriptsize
    \centering
    \vspace{-01.61800em}
    \caption{A summary of limitations in recent works.}
    \vspace{-01.2em}
    \setlength{\tabcolsep}{1.4pt}
    \begin{tabular}{m{3.5cm}|l|l|l}
    \hline
        \cellcolor [gray]{0.85}\textbf{Model name}  & \cellcolor [gray]{0.85}\textbf{BWE} & \cellcolor [gray]{0.85}\textbf{Single-Modal SE} & \cellcolor [gray]{0.85}\textbf{Multi-modal SE} \\ 
        \hline
        \hline
        ATS-UNet \cite{li2022enabling}, TFiLM \cite{birnbaum2019temporal}, AFiLM \cite{rakotonirina2021self}, TRAMBA \cite{sui2024tramba}, AERO \cite{mandel2023aero}, EBEN \cite{hauret2023eben}, HiFi++ \cite{kim2023hifi++}, NVSR \cite{liu2022neural}, NU-Wave \cite{lee2021nu}  & \checkmark &  \checkmark & \xmark  \\ 
        \hline
         ClearSpeech \cite{ma2023clearspeech}, VibVoice \cite{he2023vibvoice}  & \xmark &  \xmark &  \checkmark  \\ 
        \hline
         SEANet \cite{tagliasacchi2020seanet}  & \xmark & \checkmark  &  \checkmark  \\ 
         \hline
         \hline
        \textbf{Proposed} & \textbf{\checkmark} &  \textbf{\checkmark} &  \textbf{\checkmark}  \\ 
        \hline
    \end{tabular}
    \vspace{-01.60em}
    \label{table:summaryjointBWE}
\end{table}

\vspace{-0.820em}
\section{Preliminary}
\label{sec:preliminary}
\vspace{-0.40em}

\subsection{Power at Sub-Nyquist  Frequencies and  Bit Resolutions}
\label{subsec:Power at different sub-Nyquist sampling }
\vspace{-0.40em}

The power consumption $P$ of ADCs in hearables increases with sampling rates and resolutions, following $P = k \cdot f_s \cdot 2^N$, where  $k$ is a proportionality constant, $N$ is the bit resolution, and $f_s$ is the sampling frequency of ADCs. Traditional methods require ADCs to operate at high sampling frequencies (i.e., $>$16 kHz) and bit resolution (i.e., 12-24 bits) to accurately capture wideband audio in hearables. However, lowering sampling frequencies and bit resolution offers a powerful approach to reducing energy consumption in low-power applications like hearables. 

To support this claim, we conduct experiments with an ACM (i.e., part $\#$ B\&K Type 4192 \cite{bk4192_datasheet}) and an in-built ADC of NRF52840 \cite{nordic_nrf52840}. We get a relative power savings of 2.45x between \{16 kHz, 12-bits\} vs \{4 kHz, 8 bits\} computations. The power savings will be even greater for high-fidelity hearables, which use 22 kHz and 24-bit resolutions. \textit{\ul{CAPS considers the impact of the low bit resolution in BWE algorithms to save power in hearables that is completely absent in SOTA research.}}

\vspace{-0.73em}
\section{Architecture Design}
\label{sec:CAPS ARCHITECTURE DESIGN}
\vspace{-0.3em}

CAPS is engineered to achieve the following objectives: 
\vspace{-0.3em}

\begin{itemize}

\item CAPS will enable streaming enhancement and low-power and low-memory solutions that will make CAPS deployable on mobile platforms. 


\item CAPS will handle dynamic interferences from extremely noisy conditions in mobile scenarios using joint BWE and multimodal SE algorithms and reconstruct audio from low sampling frequency and bit resolutions.

\end{itemize}
\vspace{-0.3em}

CAPS adopts a cascaded model (Fig. \ref{fig:basic_model} (Middle \& Right)).

\vspace{-0.60em}
\subsection{Spectral Enhancement Network (SEN)}
\label{subsec:Spectral enhancement network}
\vspace{-0.3em}

The low-resolution (e.g., sampled at 4 kHz) and noisy audio from the ACM is first converted to a 2D T-F spectrogram and is given as input to the spectral enhancement network (SEN) (see Fig. \ref{fig:basic_model}). The SEN is designed to extract features from the spectrogram at the frame level and convert it to its high-resolution version. The idea is to simplify the task for the remaining part of CAPS that should transform this 2D representation to a 1D sequence for waveform-based processing.

The SEN is a 2D convolutional U-Net framework with 5 residual layers as encoders (E), 5 residual layers as decoders (D), and a Mamba \cite{gu2023mamba} block as the bottleneck. Each residual layer includes batch normalization followed by leaky ReLU activations (see Table \ref{table:spectral enhancement network} for details). Dilations are used to enlarge the receptive fields, capturing local features over a dilated window and modeling inter-phoneme dependencies in audio signals. The Mamba block in the bottleneck captures global correlations among consecutive phonemes from the spectrogram. Because Mamba is a sequence modeling architecture originally designed for 1D sequences, we adapt it to the 2D bottleneck by flattening the spatial dimensions into a sequence, applying Mamba, and then reshaping back. Moreover, Mamba has linear-time inference complexity compared to the quadratic complexity of Transformers, making it well suited for fast inference in low-power applications such as hearables.


\begin{table}[ht!]
\vspace{-01.21800em}
\scriptsize
\setlength{\tabcolsep}{0.8pt}
    \centering
    \caption{The details of the SEN and APEN.}
    \vspace{-01.31800em}
    \begin{tabular}{l | l|l|l|l | l|l|l|l | l|l||l|l}
     \hline

              & \multicolumn{10}{c||} {\cellcolor [gray]{0.85}\textbf{SEN}} &  \multicolumn{2}{c} {\cellcolor [gray]{0.85}\textbf{APEN}} \\ 
    \hline
        \cellcolor [gray]{0.85}\textbf{Layer} & \cellcolor [gray]{0.85}\textbf{E1} & \cellcolor [gray]{0.85}\textbf{E2} & \cellcolor [gray]{0.85}\textbf{E3} & \cellcolor [gray]{0.85}\textbf{E4} & \cellcolor [gray]{0.85}\textbf{E5} & \cellcolor [gray]{0.85}\textbf{D1} & \cellcolor [gray]{0.85}\textbf{D2} & \cellcolor [gray]{0.85}\textbf{D3} & \cellcolor [gray]{0.85}\textbf{D4} & \cellcolor [gray]{0.85}\textbf{D5} & \cellcolor [gray]{0.85}\textbf{Conv1D} & \cellcolor [gray]{0.85}\textbf{DWConv1D}\\ 
        \hline
        \hline
        Kernels & 8 &  16 & 24 & 32 & 64 & 64 & 32 & 24 & 16 & 8 & 512 & 512  \\ 
        \hline
       Kernel size & 4x4 & 4x4 & 4x4 & 4x4 & 4x4 & 4x4 & 4x4 & 4x4 & 4x4 & 4x4 & 7x1 & 7x1\\
       \hline
        Dilation & 1 & 1  & 2 & 3 & 5  &  & & & & & 1 & 1 \\ 
        \hline
    \end{tabular}
    \vspace{-01.20em}
    \label{table:spectral enhancement network}
\end{table}


\vspace{-0.72em}
\subsection{Upsampling Network (UN)}
\label{subsec:Upsampling network}
\vspace{-0.3em}

The upsampling network (UN) converts the spectrum to its equivalent waveform-based signals and is inspired by version 2 of HiFi-GAN \cite{kong2020hifi}. 
The UN has multiple upsampling layers, which have transposed convolutions with kernel size twice the stride, achieving an overall 256× increase in resolution through 4 stages (8×, 8×, 2×, and 2×). Each upsampling layer is followed by a residual block comprising 3 dilated convolutions with dilation rates of 1, 3, and 9 and kernel size 3, yielding an effective receptive field of 27 timesteps. Stacking dilated convolutions expands the receptive field exponentially, enabling the model to capture long-range temporal dependencies. Leaky ReLU is applied after each dilated convolution for stable training.

\vspace{-0.8520em}
\subsection{Amplitude-Phase Enhancement Network (APEN)}
\label{subsec:Spectral-phase enhancement network}
\vspace{-0.520em}

Amplitude-Phase Enhancement Network (APEN) is designed to fuse features from both the acoustic (i.e., ACM) and vibration (i.e., BCM) modalities and to remove artifacts and noise in both the \textit{amplitude and phase domains} from the output waveform in a learnable way. \textit{As signals from BCMs are less noisy compared to the signals from ACMs, the enhanced signals of ACMs from the SEN are further improved using the less noisy signal of BCMs by the APEN.} Specifically, the 1D sequences from the upsampling network are concatenated with the noisy 1D waveform from BCMs. 
This module helps to generate a clean phase from the noisy phase for the successful reconstruction of audio signals in extremely noisy conditions (see Fig. \ref{fig:basic_model}).


The waveform from the UN is first processed by a short-time Fourier transform (STFT), producing an amplitude spectrum $X_a \in \mathbb{R}^{T\times F}$ and a wrapped phase spectrum $X_p \in \mathbb{R}^{T\times F}$, where $T=125$ and $F=513$ denote the numbers of temporal frames and frequency bins, respectively. The amplitude $X_a$ and phase $X_p$ streams use identical networks consisting of a series of 1D convolution operations with mutual coupling between the two streams. These 1D convolutions comprise a cascade of a large-kernel depth-wise convolutional layer and a pair of point-wise convolutional layers that expand and then restore the feature dimensions. The depth-wise (DW) convolution is implemented with Conv1D, and the point-wise convolution is implemented with linear layers (see Table \ref{table:spectral enhancement network} for details). Layer normalization \cite{ba2016layer} and GELU activation \cite{hendrycks2016gaussian} are interleaved between layers. Finally, a residual connection is added before the output to prevent gradient vanishing.



\vspace{-0.700em}
\subsection{Loss Functions}
\label{subsec:Combined loss in the frequency, time, phase domains}
\vspace{-0.300em}

\textbf{Multi-period loss:} 
CAPS reshapes the reconstructed 1D waveform into a 2D representation by segmenting it based on a specific period (p) of 5 and 7 samples, denoted by $P^R_{5}$ and $P^R_{7}$, respectively. After reshaping, the dimension of the 2D tensor is (1, T / p, p), where T = number of audio samples. CAPS also generate $P^C_{5}$ and $P^C_{7}$ from clean audio for the same p = \{5, 7\} and calculates MAE for 2D tensors that is known as multi-period loss = $|\sum_{x,y}^{} P^C_5 - \sum_{x,y}^{} P^R_5| + |\sum_{x,y}^{} P^C_7 - \sum_{x,y}^{} P^Rs_7|$. CAPS uses only 2 periods (i.e., p = 5 and 7) compared to 5 periods and 6 convolution layers in \cite{kong2020hifi} to keep the network size small without sacrificing audio quality. 

\textbf{Phase-spectrum loss function:} 
Considering the phase wrapping issue \cite{ai2023neural}, CAPS proposes instantaneous phase and group delay anti-wrapping loss as $\frac{1}{TF} \sum_{}^{TF} |f_{AW}(Y^C_P- Y^R_P)|$ and $\frac{1}{TF} \sum_{}^{TF} |f_{AW}(Y^C_{GD}- Y^R_{GD})|$, respectively. Here, \{$Y^C_P$, $Y^C_{GD}$\} and \{$Y^R_P$, $Y^R_{GD}$\}
are the instantaneous phase and group delay for the clean and reconstructed audio, respectively. The group delay $Y^C_{GD}$ and $Y^E_{GD}$ are calculated by taking differentiation along the frequency axis of the T-F spectrogram. The $f_{AW}(x)$ denotes the anti-wrapping function, which is defined as: $f_{AW}(x) = |x - 2\pi\cdot \text{round}(\frac{x}{2\pi})|, x \in \mathbb{R}$. 

\textbf{Multi-scale loss:} Inspired by MelGAN \cite{kumar2019melgan}, CAPS uses a multi-scale (MS) loss where mean absolute error (MAE) is calculated among clean and reconstructed audio for three downsampling ratio - 1x, 2x, and 4x, expressed by $\frac{1}{3} \sum_{}^{} |D^C_1 - D^R_1| +  |D^C_2 - D^R_2| + |D^C_3 - D^R_3|$, where $D^C_n$ and $D^R_n$ denotes clean and reconstructed audio and $n$=3.






\vspace{-0.95em}
\section{Implementation}
\label{sec:implementation}
\vspace{-0.5em}

\subsection{Wearable Platform and Mobile Platform Design}
\label{subsec:Hardware design}
\vspace{-0.53em}

\begin{figure}[ht!]
\vspace{-0.95em}
    \centering
    \includegraphics[width=0.35\textwidth,height=0.1\textheight]{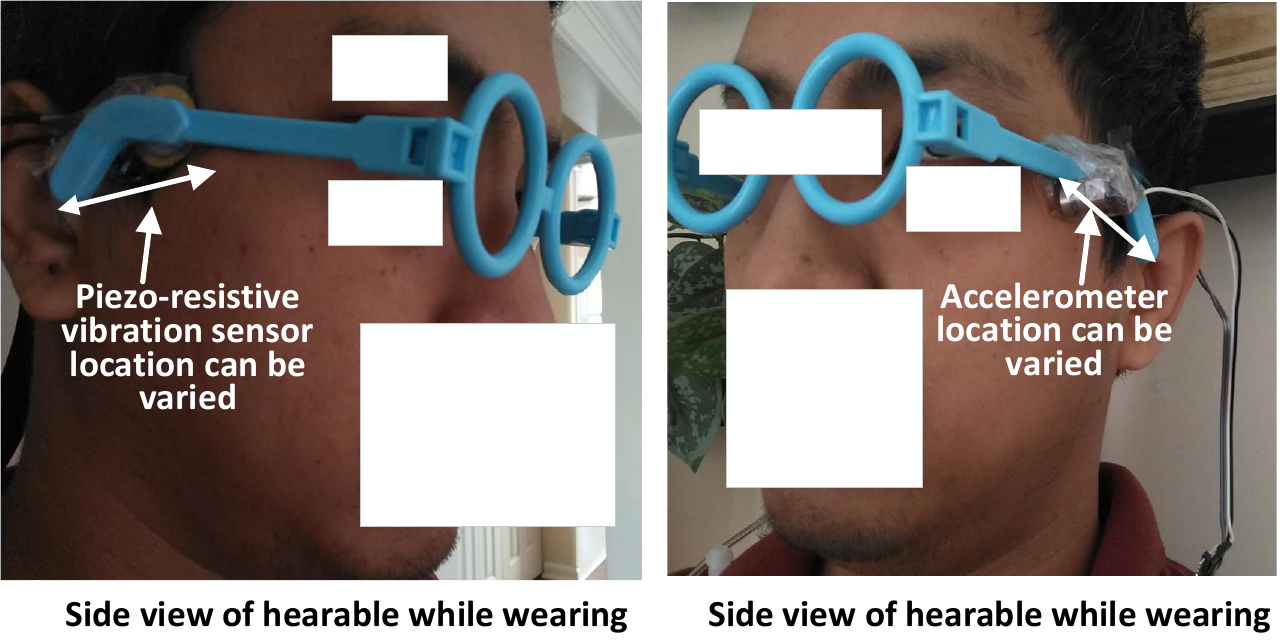}
    \vspace{-01.20em}
    \caption{A prototype of hearable.}
    \label{fig:hardwaresetup}
    \vspace{-1.6em}
\end{figure}

We use  a piezo-resistive vibration sensor (part\# CEB-27032-L100) \cite{ceb27032l100} and  an accelerometer (part\# 352C33) \cite{pcb352c33} as BCMs, placed on an off-the-shelf frame for collecting vibration near the earbone (see Fig. \ref{fig:hardwaresetup}). We use the built-in ADC from the NRF52840 chip to sample the signals from the BCMs and transmit them over Bluetooth to a mobile platform. This helps us to vary the sampling frequency from 4 kHz to 22 kHz and ADC bit resolution (8 to 12 bits) while sampling from BCMs.

As our low-resolution audio sampled from BCMs will go through joint BWE and multimodal SE in a mobile platform, we choose Google Pixel7 having the Google Tensor G2 chipset and Mali-G710 MP7 GPUs as the mobile platform. 


\vspace{-0.85em}
\subsection{Dataset Collection}
\label{subsec:Dataset collection}
\vspace{-0.4em}


Although there are a few datasets \cite{hauret2024vibravox, wang2022end} available for multimodal SE, they don't have simultaneous data from both vibration and accelerometer sensors with microphones at different ADC bit resolutions. Therefore, we ask 20 persons (11 males and 9 females, age: 18-36) for data collection for 45 minutes from the ACM and the two BCMs simultaneously at a 22 kHz sampling frequency  for 3 different bit resolutions - 12, 10, and 8 bits. We select text from the VCTK dataset \cite{yamagishi2019cstr}, which contains a variety of dialogues. For this purpose, we convert the VCTK audio to text using Whisper \cite{whisper2022}.  We apply a high-pass filter with a cut-off frequency of 5 Hz to remove any body movement. 

Two noise sources - non-speech and speech noise - in -7 to 5 dB are used. For non-speech noises, we choose from a diverse set of noise types from \cite{font2013freesound}. For speech noises,  we employ speech samples from Librispeech from different speakers.




\vspace{-0.70em}
\subsection{Model Training}
\label{subsec:Model training}
\vspace{-0.30em}

The key training parameters include a batch size of 8 with $\sim$50 epochs, and the Adam optimizer with a learning rate of \(1\times10^{-4}\), weight decay of \(1\times10^{-5}\), and momentum parameters \(\beta_1=0.5\) and \(\beta_2=0.999\). The learning rate is scheduled using cosine annealing warm restarts (with \(T_0=10\) and \(T_{\text{mult}}=1\)), gradient clipping (max norm of 10), and gradient accumulation (over 2 batches) to ensure stability. Training is first executed on a desktop with a 4090 GPU with Intel(R) Silver 4310 CPU.

\textbf{Preparing to evaluate on Google Pixel7:} 
To evaluate the model on the Google Pixel7, we perform several additional steps. Training is first conducted in PyTorch on the desktop. The trained PyTorch model is then exported to ONNX and converted to TensorFlow Lite (TFLite) via ONNX-TensorFlow \cite{onnx_tf} on the desktop. Next, the TFLite model is integrated into the TensorFlow Lite Android API. Finally, we run the TFLite model using the TFLite GPU Delegate on the Google Pixel7 to leverage the Pixel’s Mali GPU.

\vspace{-0.90em}
\subsection{Evaluation Metrics}
\label{subsec:Evaluation Metrics}
\vspace{-0.40em}

To comprehensively evaluate in terms of intelligibility, fidelity, and perceived quality, we use Log-Spectral Distance (LSD), Short-Time Objective Intelligibility (STOI), Perceptual Evaluation of Speech Quality (PESQ), Scale-Invariant Signal-to-Distortion Ratio (SI-SDR), Non-Intrusive Speech Quality Assessment - Mean Opinion Score (NISQA-MOS), and Virtual Speech Quality Objective Listener (VISQOL).

\vspace{-0.90em}
\subsection{Base Models}
\label{subsec:Base models}
\vspace{-0.4em}

We compare six base models. Two (TFiLM, VibVoice) of the models are pure U-Net based, and four (SEANet, AERO, EBEN, HiFi++) are pure GAN based. 

\vspace{-0.70em}
\section{Performance Evaluation}
\label{sec:performance evaluation}



\vspace{-0.40em}
\subsection{Real-Time BWE and Multimodel SE with Inference}
\label{subsec:Evaluation for joint BWE and multi-modal SE}
\vspace{-0.40em}


Please note that VibVoice and SEANet are multimodal SE models by default, whereas TFiLM, AERO, EBEN, and HiFi++ are single-modal SE models. To compare CAPS with the single-modal SE models, we convert our multimodal CAPS into a single-modal model by using a single-input network (i.e., we do not feed the BCM to the APEN) and name this version CAPS-single. To diversify the evaluation, we introduce the music dataset MagnaTagATune \cite{law2009evaluation}. The results are shown in Table \ref{table:evaluationforSE} for 4–16 kHz BWE with noisy data at 12-bit ADC resolution for the vibration sensor on \textit{the desktop only.}

\begin{table}[ht!]
\vspace{-01.00em}
    \scriptsize
\setlength{\tabcolsep}{0.8pt}
\renewcommand{\arraystretch}{0.8}
    \centering
    \caption{BWE and multimodal SE for 12-bit resolutions on the desktop for 4 - 16 kHz upsampling with noisy data. Here, L = LSD, V = VISQOL, N = NISQA-MOS, S = SI-SDR, P = PESQ, ST = STOI, Param = Parameter and Infer. = Inference.}
    \vspace{-01.41800em}
    \begin{tabular}
     {m{01.25cm}|m{0.7 cm}|m{0.6 cm}|m{0.6 cm}|c|m{0.5 cm}|c|c|c|c|c|c}
    \hline

         \cellcolor [gray]{0.85}\textbf{Model} & \cellcolor [gray]{0.85}\textbf{FLOPs (G)} & \cellcolor [gray]{0.85}\textbf{Param (M)} & \cellcolor [gray]{0.85}\textbf{Size (MB)} & \cellcolor [gray]{0.85}\textbf{Dataset}  & \cellcolor [gray]{0.85}\textbf{Infer. (ms)}  & \cellcolor [gray]{0.85}\textbf{L $\downarrow$} & \cellcolor [gray]{0.85}\textbf{V $\uparrow$} & \cellcolor [gray]{0.85}\textbf{N $\uparrow$} & \cellcolor [gray]{0.85}\textbf{S $\uparrow$} & \cellcolor [gray]{0.85}\textbf{P $\uparrow$} & \cellcolor [gray]{0.85}\textbf{ST $\uparrow$} \\ 
         \hline
         
        \hline
        \hline
         \multirow{2}{*} {Unprocessed}  &  & &  & VCTK &   &  2.78 & 1.84  & 1.27 & 8.54 & 1.11 & 0.79 \\
         &  & &  & Magna &   &  2.85 &  1.62  & 1.21 & 7.28 & 1.03 & 0.78  \\
        
        \Xhline{3\arrayrulewidth}

         \multirow{2}{*} {TFiLM \cite{birnbaum2019temporal}} & \multirow{2}{*} {234.7} & \multirow{2}{*} {68.2} & \multirow{2}{*} {260.3} & VCTK & 4.85  & 1.68    &  3.73    &  3.53   &  10.28   &  2.03 &  0.81  \\ 
          & &  &  & Magna &  4.85 &  1.70   &  3.67    & 3.46    &  9.41    &  1.99 &  0.81   \\ 
\hline

\multirow{2}{*} {VibVoice \cite{he2023vibvoice}} & \multirow{2}{*} {79.4} & \multirow{2}{*} {23.14} & \multirow{2}{*} {5.8 }& VCTK & 17.2  & 3.1  &  2.73    & 2.51    &  12.48    &  2.05 &  0.81  \\ 
         & &  &   & Magna & 17.2  & 3.28   &  2.66    & 2.43    &   11.21  &  2.01  &   0.80   \\ 

        \Xhline{3\arrayrulewidth}   
          
         \multirow{2}{*} {AERO \cite{mandel2023aero}} & \multirow{2}{*} {124.8} & \multirow{2}{*} {36.3} & \multirow{2}{*} {138.7} &  VCTK & 36 &  0.97  &  4.16   &  4.03    &  17.03 &  2.93   &   0.89  \\ 
         & &  &  & Magna &  36 & 0.99    &  4.10   &  3.98    &   16.29  &  2.89  &   0.88  \\ 
          
         \hline
        \multirow{2}{*} {EBEN \cite{hauret2023eben}} & \multirow{2}{*} {101.4} & \multirow{2}{*} {29.7} & \multirow{2}{*} {113.3} & VCTK  & 12.7 &  1.15 &   3.78   & 3.65    & 14.23  & 2.57  & 0.84  \\ 
         & & &  &  Magna  & 12.7 & 1.17    &   3.76    & 3.63    & 13.13  & 2.54 & 0.88   \\ 
           
         \hline
        \multirow{2}{*} {HiFi++ \cite{kim2023hifi++}} & \multirow{2}{*} {250.5} & \multirow{2}{*} {72.2} & \multirow{2}{*} {259.9} &  VCTK  &  6.3 & 0.89   &   4.18   &  4.11   & 17.48   &  2.85 &  0.90  \\ 
         & & & &  Magna &  6.3 & 0.91   &   4.13   &  4.07   & 16.65   & 2.83   &   0.89  \\ 
          
         \hline
         \multirow{2}{*} {SEANet \cite{tagliasacchi2020seanet}}   &  \multirow{2}{*} {225.1} & \multirow{2}{*} {64.9} & \multirow{2}{*} {240.1} & VCTK & 9.18 & 1.39  &   3.89  &   3.78   &  14.31  & 2.43 &   0.89  \\
         & & & &  Magna  & 9.18 & 1.44   &  3.79    &   3.67    & 13.74     & 2.40 &  0.87   \\
         
        \hline
        \hline
       \multirow{2}{*} {\textbf{CAPS}}  &  \multirow{2}{*} {10.01} & \multirow{2}{*} {2.85}  & \multirow{2}{*} {11.04} & VCTK & \textbf{1.36} & \textbf{0.87} &  \textbf{4.15}  &   \textbf{4.13} &   \textbf{16.99}    &  \textbf{2.99} & \textbf{0.90}  \\
       & & &  &  Magna &  \textbf{1.36}  & \textbf{0.88} &  \textbf{4.11}  &   \textbf{4.10} &   \textbf{16.61}  &  \textbf{2.92}  &  \textbf{0.90} \\

          \multirow{2}{*} {\vspace{3mm}\textbf{CAPS-}} & \multirow{2}{*} {10.01} & \multirow{2}{*} {2.85} & \multirow{2}{*} {11.04} &  VCTK & \textbf{1.36}  & \textbf{0.87} &  \textbf{4.10} &   \textbf{4.11} &   \textbf{16.97}    &  \textbf{2.97} & \textbf{0.90}   \\
       \textbf{single} & &  & &  Magna &  \textbf{1.36}  & \textbf{0.88} &  \textbf{4.09}  &   \textbf{4.09} &   \textbf{16.71}  &  \textbf{2.95}  &  \textbf{0.90} \\
        \hline
    \end{tabular}
    \vspace{-0.20em}
    \label{table:evaluationforSE}
    \vspace{-01.400em}
\end{table}

Table \ref{table:evaluationforSE} shows that CAPS achieves better perceptual quality and intelligibility than SOTA U-Net and GAN models on both speech and music datasets, while having the lowest inference time (i.e., 1.36 ms), which is 4.63x lower than HiFi++ and 27x lower than AERO (i.e., the two best-performing models). Please note that CAPS is also the smallest model, with 25x fewer parameters than HiFi++ and 13x fewer parameters than AERO.

Table \ref{table:evaluationforSEforvibration and accel} shows the real-time performance of CAPS on both desktop and Google Pixel7 platforms.
We keep the model quantization the same for both desktop and Pixel7 while evaluating the inference time. All the models in Table \ref{table:evaluationforSEforvibration and accel} are suitable for real-time operation on the desktop, as the inference time is lower than the audio frame (1s). 
Moreover, for the streaming operation, one-way delays up to 150 ms are considered acceptable for most user applications, including voice calls, according to the recommendation of the International Telecommunication Union (ITU) G.114 \cite{ITU-G114}. 
\textit{Therefore, our proposed CAPS is the only model which has inference time smaller than 150 ms on both desktop (1.36 ms) and Pixel7 (55.11 ms), indicating its capability of streaming on both desktop and mobile platforms.}

\begin{table}[ht!]
\vspace{-0.81800em}
    \scriptsize
\setlength{\tabcolsep}{0.6pt}
\renewcommand{\arraystretch}{0.8}
    \centering
    \caption{BWE and SE for 12-bit resolutions on desktop and Google Pixel7 for 4-16 kHz upsampling with noisy data for both vibration sensor and accelerometer.}
    \vspace{-01.381800em}
    \begin{tabular}
     {l|l|l|l|l|l|l|l||l|l|l|l|l}
    \hline

         \cellcolor [gray]{0.85}\textbf{Model}  & \cellcolor [gray]{0.85}\textbf{Device} & \cellcolor [gray]{0.85}\textbf{Infer.} & \multicolumn{5}{c||} {\cellcolor [gray]{0.85}\textbf{Vibration sensor}} &  \multicolumn{5}{|c} {\cellcolor [gray]{0.85}\textbf{Accelerometer}} \\ 
         \hline
          &  & (ms)  &  \textbf{L$\downarrow$} & \textbf{V$\uparrow$}  & \textbf{N$\uparrow$} & \textbf{S$\uparrow$}  & \textbf{P$\uparrow$}  & \textbf{L$\downarrow$} & \textbf{V$\uparrow$}  & \textbf{N$\uparrow$} & \textbf{S$\uparrow$}  & \textbf{P$\uparrow$} \\
         
        \hline
        \hline
        

         \multirow{2}{*} {\vspace{3mm}TFiLM\cite{birnbaum2019temporal}}   & Desktop & 4.85 & 1.68   &  3.73   &  3.53   &  10.28   &  2.03      & 1.82   &  3.47    & 3.36    &  9.58    &  1.91   \\ 
           (U-Net)  & Pixel7  & 198 & 1.69   &  3.70     & 3.50   &  10.19   &  2.00    &  1.83  &  3.45   & 3.34   &   9.52  &  1.88    \\ 
\hline

\multirow{2}{*} {\vspace{3mm}VibVoice\cite{he2023vibvoice}}  & Desktop & 17.2  & 3.1  &  2.73    & 2.51    &  12.48    &  2.05   &  3.5   &  2.54    & 2.41    &  11.43    &  1.93  \\ 
          (U-Net) &  Pixel7  & 707  & 3.11   &  2.70    & 2.47    &   12.45  &  2.01    &  3.48   &  2.51    & 2.38   &   11.39  &  1.90    \\ 

        \Xhline{3\arrayrulewidth}   
          
         \multirow{2}{*} {\vspace{3mm}AERO\cite{mandel2023aero}}  & Desktop & 36 & 0.97  &  4.16   &  4.03    &  17.03 &  2.93    &  1.03   &  4.04   &  3.91    &  16.65 &  3.86   \\ 
          (GAN)  & Pixel7  & 1441 &  0.98    &  4.14   &  4.01    &   16.97  &  2.89   &  1.05    &  4.01   &  3.89    &   16.57  &  3.84    \\ 
          
         \hline
        \multirow{2}{*} {\vspace{3mm}EBEN\cite{hauret2023eben}}  & Desktop & 12.7& 1.15 &   3.78   & 3.65    & 14.23  & 2.57   &  1.21     &   3.58  & 3.37    & 13.27  & 2.39   \\ 
         (GAN)   & Pixel7  & 520 & 1.17    &   3.76    & 3.63    & 14.19  & 2.56  &  1.22   &   3.54    &  3.33   & 13.21  & 2.35  \\ 
           
         \hline
        \multirow{2}{*} {\vspace{3mm}HiFi++\cite{kim2023hifi++}}  & Desktop & 6.3 & 0.89   &   4.18  & 4.11   & 17.48   &  2.85  &   0.92   &   4.10   &  4.03   & 16.87   & 2.81 \\ 
         (GAN) & Pixel7  & 258 & 0.91   &   4.15   &  4.09   & 17.43   & 2.83  &  0.94   &   4.07   &  3.98   & 16.71   & 2.78   \\ 
          
         
        \hline
        \hline
       \multirow{2}{*} {\textbf{CAPS}}   & Desktop & \textbf{1.36} & \textbf{0.87} &  \textbf{4.15}  &   \textbf{4.13} &   \textbf{16.99}    &  \textbf{2.99} &  \textbf{0.88} &  \textbf{4.12}  &   \textbf{4.10} &   \textbf{16.65}    &  \textbf {2.90}  \\
           & Pixel7  & \textbf{55.11} & \textbf{0.88} &  \textbf{4.13}  &   \textbf{4.15} &   \textbf{17.21}  &  \textbf{2.97}  &  \textbf{0.89} &  \textbf{4.11}  &   \textbf{4.09} &   \textbf{16.78}  &  \textbf{2.89}   \\

          \multirow{2}{*} {\vspace{3mm}\textbf{CAPS-}}   & Desktop &  \textbf{1.36} & \textbf{0.87} &  \textbf{4.10}  &   \textbf{4.11} &   \textbf{16.97}    &  \textbf{2.97}  &  \textbf{0.88} &  \textbf{4.09}  &   \textbf{4.10} &   \textbf{16.87}    &  \textbf{2.89}   \\
           \textbf{single} & Pixel7  & \textbf{55.11} & \textbf{0.88} &  \textbf{4.10}  &   \textbf{4.12} &   \textbf{17.11}  &  \textbf{2.95}  &  \textbf{0.90} &  \textbf{4.09}  &   \textbf{4.07} &   \textbf{16.58}  &  \textbf{2.87} \\
        \hline
    \end{tabular}
    \vspace{-0.20em}
    \label{table:evaluationforSEforvibration and accel}
    \vspace{-02.5800em}
\end{table}

\vspace{-0.95em}
\subsection{Evaluation of power consumption}
\label{subsec:Evaluation for power consumptio}
\vspace{-0.5em}


We now evaluate how much power can be saved by reducing the sampling frequency and bit resolution at the hearables’ ADC, and how much additional power is required to run CAPS on mobile platforms such as smartphones. We vary the sampling frequency of the NRF52840 from 4 kHz to 24 kHz at 8-bit, 10-bit, and 12-bit resolutions, and measure power for each configuration in Table \ref{table:adcpowerbit} with EasyDMA enabled.

\begin{table}[h!]
\vspace{-01.100em}
\scriptsize
\setlength{\tabcolsep}{0.6pt}
\renewcommand{\arraystretch}{0.5}
\centering
\caption{Power at different sampling and resolutions.}
\vspace{-01.471800em}
\begin{tabular}{c|c|c|c}
\hline
 \cellcolor [gray]{0.85}\textbf{Sampling Rate (Hz)} &  \cellcolor [gray]{0.85}\textbf{Resolution (bits)} &  \cellcolor [gray]{0.85}\textbf{Current (µA)} &  \cellcolor [gray]{0.85}\textbf{Power (mW) @ 3.0 V} \\
\hline
\multirow{3}{*}{4 kHz} & 8-bit  & 234  & 0.702  \\
                       & 10-bit & 248  & 0.744  \\
                       & 12-bit & 275  & 0.825  \\
\hline
\multirow{3}{*}{8 kHz} & 8-bit  & 319  & 0.957  \\
                       & 10-bit & 338  & 1.014 \\
                       & 12-bit & 375  & 1.125 \\
\hline
\multirow{3}{*}{16 kHz} 
      & 8-bit   & 489 & 1 467 \\
      & 10-bit  & 518 & 1 554 \\
      & 12-bit  & 575 & 1 725 \\
\hline
\multirow{3}{*}{24 kHz} & 8-bit  & 659  & 1.977 \\
                        & 10-bit & 698  & 2.094 \\
                        & 12-bit & 775  & 2.325 \\
\hline
\end{tabular}
\label{table:adcpowerbit}
\vspace{-01.51800em}
\end{table}


Table \ref{table:adcpowerbit} indicates that if we use \{4 kHz, 8-bit\} sampling instead of \{24 kHz, 12-bit\} sampling in hearables, we can save 2.325/0.702 = 3.31x power in hearables. That means that we can increase the battery life by $\sim$3.31x for hearables. Therefore, the idea is that CAPS will reduce the sampling frequency and bit-resolutions in hearables from \{24 kHz, 12-bit\} or \{16 kHz, 12-bit\} to  \{4 kHz, 8-bit\} to save power in hearables. Later, CAPS will restore the low-resolution audio from \{4 kHz, 8-bit\}  to \{24 kHz, 12-bit\} or \{16 kHz, 12-bit\} on mobile platforms to provide the same audio quality.

\begin{table}[ht!]
\vspace{-01.100em}
    \scriptsize
\setlength{\tabcolsep}{0.53pt}
\renewcommand{\arraystretch}{0.5}
    \centering
    \caption{Power in Pixel7 at different transmission rates.}
    \vspace{-01.2900em}
    \begin{tabular}{ m{1.5cm}|m{2.0cm}|m{2.6cm}|l }
    \hline
         \cellcolor [gray]{0.85}\textbf{Transmission rate} & \cellcolor [gray]{0.85}\textbf{Power for CAPS only} & \cellcolor [gray]{0.85}\textbf{Power for different transmission rates only} &  \cellcolor [gray]{0.85}\textbf{Total power}\\ 
        \hline
        \hline
        64 kbps  & 1.133 W  &  3.54 mW  &  1.136 W \\
        128 kbps  & 1.141 W &   4.18 mW  & 1.145 W\\
        256 kbps  &  1.177 W &   5.98 mW   & 1.183 W\\
        \hline
                  &           &            & Avg. = 1.15 W \\
        \hline
    \end{tabular}
    \vspace{-0.20em}
    \label{table:poeringooglepixel}
    \vspace{-0.95800em}
\end{table}

\textbf{\textit{The next question is how much power is needed to restore audio using CAPS on mobile platforms.}} We evaluate CAPS on Google Pixel7 at different transmission rates from hearables to Pixel7 (see Table \ref{table:poeringooglepixel}). Power increases with transmission rate because higher rates require more computation for spectrograms and raw waveforms. The average power consumption of $\sim$1.15 W for running CAPS on smartphones such as Pixel7 is negligible compared to the 3.31x power saving in hearables, since smartphones typically have batteries about 100x larger than hearables. \textit{For example, Samsung Galaxy Buds2 Pro has a 50 mAh battery, while Pixel7 has a 4355 mAh battery. The 1.15 W power consumption by CAPS would take about 13 hours to drain Pixel7’s 4355 mAh battery, which is trivial.}


\vspace{-0.91800em}
\subsection{ADC bit resolutions and sampling frequencies}
\label{subsec:Evaluation at different bit resolutions}
\vspace{-0.500em}

We vary the bit resolution from 8 to 12 bits for 4-16 kHz and 4-22 kHz upsampling, and evaluate CAPS on a desktop. Table \ref{table:bitresolution and BWE} shows that CAPS’s performance degrades at low bit resolutions due to increased quantization error during sampling. However, CAPS still outperforms the best-performing models.

\begin{table}[ht!]
\vspace{-01.300em}
    \scriptsize
\setlength{\tabcolsep}{0.3pt}
\renewcommand{\arraystretch}{0.5}
    \centering
    \caption{ Evaluating SE for 8, 10, and 12-bit ADC resolutions on the desktop for 4 - 16 kHz and 4 - 22 kHz upsampling with noisy data for the vibration sensor.}
    \vspace{-01.400em}
    \begin{tabular}
     {m{0.7cm}|m{0.35cm}|m{01cm}|m{01.10cm}|m{01.3cm}|m{01.3cm}|m{01.0cm}|m{01.0cm}}
    \hline

         \cellcolor [gray]{0.85}\textbf{Model} & \cellcolor [gray]{0.85}\textbf{Bit}   & \cellcolor [gray]{0.85}\textbf{LSD$\downarrow$ (4-16k/4-22k)} & \cellcolor [gray]{0.85}\textbf{VISQOL \hspace{0.6cm}(4-16k/4-22k)} & \cellcolor [gray]{0.85}\textbf{NISQA-MOS  \hspace{0.0cm}(4-16 k/ 4-22k)} & \cellcolor [gray]{0.85}\textbf{SI-SDR  \hspace{0.6cm}(4-16k/4-22k)} & \cellcolor [gray]{0.85}\textbf{PESQ \hspace{0.6cm} (4-16k/4-22k)} & \cellcolor [gray]{0.85}\textbf{STOI  \hspace{0.6cm} (4-16k/4-22k)} \\ 
         \hline
         
        \hline
        \hline

         \multirow{3}{*} {AERO}  &  8  &  0.99/1.04 &  4.01/3.96   &  3.91/3.80   &  16.03/16.42 &  2.85/2.69  &  0.88/0.87    \\ 
           & 10  & 0.98/1.03    &  4.10/4.06   &  3.98/3.86    &   16.69/17.14  &  2.89/2.73 & 0.89/0.88    \\ 
          & 12  & 0.97/1.02    &  4.16/4.11   &  4.03/3.90   &   17.03/17.49  &  2.93/2.77 & 0.89/0.88    \\ 
       
         \hline
        \multirow{3}{*} {HiFi++}  &  8   & 0.91/93   &   4.01/3.96   &  3.97/3.93   & 16.28/16.14   &  2.81/2.80 &  0.89/0.88  \\ 
         &   10  & 0.90/0.92   &   4.13/3.99   &  4.07/3.97   & 16.95/16.85   & 2.83/2.82   &   0.89/0.88  \\ 
          &   12  & 0.89/0.91   &   4.18/4.05   &  4.11/4.01   & 17.48/17.29   & 2.85/2.84   &   0.90/0.89  \\

        \hline
        \hline
       \multirow{3}{*} {\textbf{CAPS}}   &  8   & \textbf{0.88/0.89} &  \textbf{4.09/3.95}  &   \textbf{4.05/3.99} &   \textbf{16.84/16.71}    &  \textbf{2.93/2.91} & \textbf{0.90/0.89}  \\
       &   10   & \textbf{0.87/0.88} &  \textbf{4.11/3.99}  &   \textbf{4.09/3.97} &   \textbf{16.89/16.77}  &  \textbf{2.96/2.95}  &  \textbf{0.90/0.90} \\

       &   12   & \textbf{0.87/0.88} &  \textbf{4.15/4.01}  &   \textbf{4.13/4.01} &   \textbf{16.99/16.75}  &  \textbf{2.99/2.97}  &  \textbf{0.90/0.90} \\

        \hline
    \end{tabular}
    \vspace{-0.20em}
    \label{table:bitresolution and BWE}
    \vspace{-01.85800em}
\end{table}

\vspace{-01.40em}
\subsection{Evaluation at other mobile platforms}
\label{subsec:Evaluation at other mobile platforms}
\vspace{-0.50em}

To generalize, we evaluate CAPS on Samsung Galaxy S21 also (see Table \ref{table:Samsung vs pixel7}). The inference speed on the S21 is 1.51x slower than Pixel7, because of Pixel7's custom tensor processing unit (TPU) in its Tensor G2 chipset. The evaluation metrics are similar to Pixel7, as both the Pixel7 and Galaxy S21 use the same CAPS model with the same quantization (float 32).

\begin{table}[ht!]
\vspace{-01.00em}
   \scriptsize
\setlength{\tabcolsep}{0.4pt}
    \centering
    \caption{Evaluation at two different mobile platforms.}
    \vspace{-01.2951800em}
    \begin{tabular}
     {c|c|c|c|c|c|c|c}
    \hline

         \cellcolor [gray]{0.85}\textbf{Model} & \cellcolor [gray]{0.85}\textbf{Platform}  & \cellcolor [gray]{0.85}\textbf{ADC resolution} & \cellcolor [gray]{0.85}\textbf{Inference}   &  \cellcolor [gray]{0.85}\textbf{Avg. Power} & \cellcolor [gray]{0.85}\textbf{L$\downarrow$} & \cellcolor [gray]{0.85}\textbf{P $\uparrow$} & \cellcolor [gray]{0.85}\textbf{ST $\uparrow$} \\ 
         \hline
         
        \hline
        \hline

        \multirow{2}{*} {\textbf{CAPS}}  &   Pixel7   & 12-bit &  55.11 ms  &   1.15 W &     0.84   &  2.99  &  0.90  \\

       &   Galaxy S21   & 12-bit &  84.3 ms  &   1.22 W &   0.84  & 2.96 & 0.90 \\

        \hline
    \end{tabular}
    \vspace{-0.20em}
    \label{table:Samsung vs pixel7}
    \vspace{-01.5800em}
\end{table}

\vspace{-0.6100em}
\subsection{Ablation study on the model components}
\label{subsec:Ablation study on the model}
\vspace{-0.400em}

Table \ref{table:Ablation study on the model} shows that replacing Mamba in the SEN yields similar performance metrics. However, the training time increases significantly from 212 s to 369 s per epoch, and the number of parameters increases slightly from 2.85 M to 2.98 M. Therefore,  we keep Mamba in our design as Mamba provides similar performance with a shorter training time and smaller parameters. We can see from Table \ref{table:Ablation study on the model} that the APEN is an important component, as its absence notably degrades the model performance for all metrics. The reason behind this is that the APEN improves the low-resolution spectral and phase information in noisy conditions. Moreover, the multi-period loss function is critical as it works in the time domain and facilitates learning low-resolution to high-resolution mapping effectively.

\begin{table}[ht!]
\vspace{-01.00em}
    \scriptsize
\setlength{\tabcolsep}{1.10pt}
    \centering
    \caption{Ablation study for 4 - 16 kHz and 12-bit resolutions.}
    \vspace{-01.3971800em}
    \begin{tabular}{ m{2.5cm}|l|l|l|l| m{1.4cm} | m{1.4cm} }
    \hline
         \cellcolor [gray]{0.85}\textbf{Method} & \cellcolor [gray]{0.85}\textbf{L $\downarrow$} & \cellcolor [gray]{0.85}\textbf{S $\uparrow$} & \cellcolor [gray]{0.85}\textbf{P $\uparrow$} & \cellcolor [gray]{0.85}\textbf{ST $\uparrow$} & \cellcolor [gray]{0.85}\textbf{Parameter} & \cellcolor [gray]{0.85}\textbf{Train time per epoch}\\ 
        \hline
        \hline
        CAPS  & 0.87  & 16.99  & 2.99  & 0.90 & 2.85 M  &  212 s     \\
        \hline
        Replace Mamba with Transformers in the SEN  & 0.85  & 17.12  & 3.03  & 0.90  & 2.98 M  &  369 s  \\
        \hline
        \hline
        without APEN  &  1.32  &  7.12 &  1.95  & 0.79 & 2.0214 M &  135 s\\
        \hline
        without multi-period loss & 0.84  & 15.94  & 2.50  & 0.81 & 3.61 M  &  212 s   \\
        \hline
    \end{tabular}
    \vspace{-0.20em}
    \label{table:Ablation study on the model}
    \vspace{-0.75800em}
\end{table}

\vspace{-0.9100em}
\subsection{Subjective Analysis}
\label{subsec:Subjective Analysis}
\vspace{-0.600em}

We conduct a mean opinion score (MOS) across 4-22 kHz extension for 8-bit resolution with 12 volunteers and get an average of 4.3 MOS on a scale from 1 (bad) to 5 (best). This provides strong evidence that CAPS consistently generates perceptually higher quality audio, favored by a wide range of listeners.

\vspace{-0.900em}
\section{Conclusion and Limitation}
\label{sec:Conclusion}
\vspace{-0.500em}

The low latency of CAPS while performing joint BWE and multimodal SE  enables streaming enhancement, low power, and low memory solutions, making CAPS deployable on mobile platforms.  Therefore, CAPS is designed to meet the high demands of smart hearables in low-power and real-world usage by bridging the gap between power and the performance of current technologies. However, this work does not consider any encryption of the sampled audio before transmission to the mobile platform from the hearables. Moreover, we do not include any codec while evaluating its performance, power, and efficiency.

\vspace{-0.600em}
\section{Generative AI Use Disclosure}

We acknowledge that we have used Elicit for finding relevant papers, and used ChatGPT for debugging codes, and finding grammatical errors

\bibliographystyle{IEEEtran}
\bibliography{mybib}

@article{sui2024tramba,
  title={Tramba: A hybrid transformer and mamba architecture for practical audio and bone conduction speech super resolution and enhancement on mobile and wearable platforms},
  author={Sui, Yueyuan and Zhao, Minghui and Xia, Junxi and Jiang, Xiaofan and Xia, Stephen},
  journal={Proceedings of the ACM on Interactive, Mobile, Wearable and Ubiquitous Technologies},
  volume={8},
  number={4},
  pages={1--29},
  year={2024},
  publisher={ACM New York, NY, USA}
}

@inproceedings{rakotonirina2021self,
  title={Self-attention for audio super-resolution},
  author={Rakotonirina, Nathana{\"e}l Carraz},
  booktitle={2021 IEEE 31st International Workshop on Machine Learning for Signal Processing (MLSP)},
  pages={1--6},
  year={2021},
  organization={IEEE}
}

@article{gu2023mamba,
  title={Mamba: Linear-time sequence modeling with selective state spaces},
  author={Gu, Albert and Dao, Tri},
  journal={arXiv preprint arXiv:2312.00752},
  year={2023}
}

@inproceedings{mandel2023aero,
  title={Aero: Audio super resolution in the spectral domain},
  author={Mandel, Moshe and Tal, Or and Adi, Yossi},
  booktitle={ICASSP 2023-2023 IEEE International Conference on Acoustics, Speech and Signal Processing (ICASSP)},
  pages={1--5},
  year={2023},
  organization={IEEE}
}

@article{ho2020denoising,
  title={Denoising diffusion probabilistic models},
  author={Ho, Jonathan and Jain, Ajay and Abbeel, Pieter},
  journal={Advances in neural information processing systems},
  volume={33},
  pages={6840--6851},
  year={2020}
}

@article{tagliasacchi2020seanet,
  title={SEANet: A multi-modal speech enhancement network},
  author={Tagliasacchi, Marco and Li, Yunpeng and Misiunas, Karolis and Roblek, Dominik},
  journal={arXiv preprint arXiv:2009.02095},
  year={2020}
}

@inproceedings{han2022nu,
  title     = {NU-Wave 2: A General Neural Audio Upsampling Model for Various Sampling Rates},
  author    = {Seungu Han and Junhyeok Lee},
  year      = {2022},
  booktitle = {Interspeech 2022},
  pages     = {4401--4405},
  doi       = {10.21437/Interspeech.2022-45},
  issn      = {2958-1796},
}

@inproceedings{liu2022neural,
  title={Neural vocoder is all you need for speech super-resolution},
  author={Liu, Haohe and Choi, Woosung and Liu, Xubo and Kong, Qiuqiang and Tian, Qiao and Wang, DeLiang},
  year      = {2022},
  booktitle = {Interspeech},

}

@inproceedings{lee2021nu,
  title     = {NU-Wave: A Diffusion Probabilistic Model for Neural Audio Upsampling},
  author    = {Junhyeok Lee and Seungu Han},
  year      = {2021},
  booktitle = {Interspeech 2021},
  pages     = {1634--1638},
  doi       = {10.21437/Interspeech.2021-36},
  issn      = {2958-1796},
}

@article{yamagishi2019cstr,
  title={CSTR VCTK Corpus: English multi-speaker corpus for CSTR voice cloning toolkit (version 0.92)},
  author={Yamagishi, Junichi and Veaux, Christophe and MacDonald, Kirsten and others},
  journal={University of Edinburgh. The Centre for Speech Technology Research (CSTR)},
  pages={271--350},
  year={2019}
}

@misc{nordic_nrf52840,
  title        = {nRF52840 Product Specification v1.1},
  author       = {Nordic Semiconductor},
  year         = {2018},
  howpublished = {\url{https://infocenter.nordicsemi.com/pdf/nRF52840\_PS\_v1.1.pdf}},
  note         = {Accessed: 2025-04-26}
}

@article{zhang2025wearse,
  title={WearSE: Enabling Streaming Speech Enhancement on Eyewear Using Acoustic Sensing},
  author={Zhang, Qian and Guo, Kaiyi and Yang, Yifei and Wang, Dong},
  journal={Proceedings of the ACM on Interactive, Mobile, Wearable and Ubiquitous Technologies},
  volume={9},
  number={1},
  pages={1--30},
  year={2025},
  publisher={ACM New York, NY, USA}
}

@article{birnbaum2019temporal,
  title={Temporal FiLM: Capturing long-range sequence dependencies with feature-wise modulations.},
  author={Birnbaum, Sawyer and Kuleshov, Volodymyr and Enam, Zayd and Koh, Pang Wei W and Ermon, Stefano},
  journal={Advances in Neural Information Processing Systems},
  volume={32},
  year={2019}
}

@misc{onnx_tf,
  author       = {ONNX-TensorFlow Community},
  title        = {ONNX-TensorFlow: Open Neural Network Exchange (ONNX) backend for TensorFlow},
  year         = {2024},
  howpublished = {\url{https://github.com/onnx/onnx-tensorflow}},
  note         = {GitHub repository}
}

@inproceedings{law2009evaluation,
  title={Evaluation of algorithms using games: The case of music tagging},
  author={Law, Edith and West, Michael and Mandel, Michael and Bay, Michael and Downie, J Stephen},
  booktitle={Proceedings of the 10th International Society for Music Information Retrieval Conference (ISMIR)},
  year={2009}
}

@article{li2022enabling,
  title={Enabling real-time on-chip audio super resolution for bone-conduction microphones},
  author={Li, Yuang and Wang, Yuntao and Liu, Xin and Shi, Yuanchun and Patel, Shwetak and Shih, Shao-Fu},
  journal={Sensors},
  volume={23},
  number={1},
  pages={35},
  year={2022},
  publisher={MDPI}
}

@misc{bk4192_datasheet,
  author       = {{Brüel \& Kjær Sound \& Vibration Measurement A/S}},
  title        = {{Type 4192: 1/2" Pressure-field Microphone – High Sensitivity}},
  year         = {2002},
  howpublished = {\url{https://www.bksv.com/en/transducers/microphones/microphone-cartridges/4192}},
  note         = {Accessed: 2025-04-10}
}

@inproceedings{kim2023hifi++,
  title={{HiFi++}: Towards Perceptually Enhanced and Computationally Efficient Neural Vocoders},
  author={Kim, Woosung and Kang, Hyeongmin and Kim, Younggun and Jung, Kyunggu and Lee, Joon Son and Lee, Sang-Hoon},
  booktitle={Proc. Interspeech},
  year={2023},
  pages={4374--4378}
}

@inproceedings{hauret2023eben,
  title={EBEN: Extreme Bandwidth Extension Network applied to speech signals captured with noise-resilient body-conduction microphones},
  author={Hauret, Julien and Joubaud, Thomas and Zimpfer, V{\'e}ronique and Bavu, {\'E}ric},
  booktitle={ICASSP 2023 - 2023 IEEE International Conference on Acoustics, Speech and Signal Processing (ICASSP)},
  pages={1--5},
  year={2023},
  organization={IEEE},
  doi={10.1109/ICASSP49357.2023.10096301}
}

@article{han2024earspeech,
  title={EarSpeech: Exploring In-Ear Occlusion Effect on Earphones for Data-efficient Airborne Speech Enhancement},
  author={Han, Feiyu and Yang, Panlong and Zuo, You and Shang, Fei and Xu, Fenglei and Li, Xiang-Yang},
  journal={Proceedings of the ACM on Interactive, Mobile, Wearable and Ubiquitous Technologies},
  volume={8},
  number={3},
  pages={104:1--104:30},
  year={2024},
  publisher={ACM},
  doi={10.1145/3678594},
  url={https://doi.org/10.1145/3678594}
}

@article{ma2023clearspeech,
  title={ClearSpeech: Improving Voice Quality of Earbuds Using Both In-Ear and Out-Ear Microphones},
  author={Ma, Dong and Dang, Ting and Ding, Ming and Balan, Rajesh Krishna},
  journal={Proceedings of the ACM on Interactive, Mobile, Wearable and Ubiquitous Technologies},
  volume={7},
  number={4},
  pages={170:1--170:25},
  year={2023},
  publisher={ACM},
  doi={10.1145/3631409}
}

@inproceedings{he2023vibvoice,
  title={VibVoice: Towards Bone-Conducted Vibration Speech Enhancement on Head-Mounted Wearables},
  author={He, Lixing and Hou, Haozheng and Shi, Shuyao and Shuai, Xian and Yan, Zhenyu},
  booktitle={Proceedings of the 21st ACM International Conference on Mobile Systems, Applications and Services},
  pages={356--369},
  year={2023},
  organization={ACM}
}

@misc{whisper2022,
  title        = {Whisper: Robust Speech Recognition via Large-Scale Weak Supervision},
  author       = {Radford, Alec and Kim, Jong Wook and Xu, Tao and Brockman, Greg and McLeavey, Christine and Sutskever, Ilya},
  year         = {2022},
  howpublished = {\url{https://github.com/openai/whisper}},
  note         = {OpenAI Technical Report}
}

@inproceedings{font2013freesound,
  title={Freesound technical demo},
  author={Font, Frederic and Roma, Gerard and Serra, Xavier},
  booktitle={Proceedings of the 21st ACM international conference on Multimedia},
  pages={411--412},
  year={2013}
}

@article{wang2022end,
  title={End-to-end multi-modal speech recognition on an air and bone conducted speech corpus},
  author={Wang, Mou and Chen, Junqi and Zhang, Xiao-Lei and Rahardja, Susanto},
  journal={IEEE/ACM Transactions on Audio, Speech, and Language Processing},
  volume={31},
  pages={513--524},
  year={2022},
  publisher={IEEE}
}

@article{hauret2024vibravox,
  title={Vibravox: A Dataset of French Speech Captured with Body-conduction Audio Sensors},
  author={Hauret, Julien and Olivier, Malo and Joubaud, Thomas and Langrenne, Christophe and Poir{\'e}e, Sarah and Zimpfer, V{\'e}ronique and Bavu, {\'E}ric},
  journal={arXiv preprint arXiv:2407.11828},
  year={2024}
}

@misc{ceb27032l100,
  author       = {{CUI Devices}},
  title        = {{CEB-27032-L100: Contact Microphone}},
  year         = {2023},
  howpublished = {\url{https://www.cuidevices.com/product/audio/speakers/contact-microphones/ceb-27032-l100}},
  note         = {Accessed: 2025-04-10},
}

@misc{pcb352c33,
  author       = {{PCB Piezotronics, Inc.}},
  title        = {{352C33: ICP® Quartz Shear Accelerometer}},
  year         = {2024},
  howpublished = {\url{https://www.pcb.com/products?m=352C33}},
  note         = {Accessed: 2025-04-10},
}

@article{hendrycks2016gaussian,
  title={Gaussian error linear units (gelus)},
  author={Hendrycks, Dan and Gimpel, Kevin},
  journal={arXiv preprint arXiv:1606.08415},
  year={2016}
}

@misc{ITU-G114,
  author       = {{International Telecommunication Union}},
  title        = {{Recommendation G.114: One-way transmission time}},
  year         = {2003},
  howpublished = {\url{https://www.itu.int/rec/T-REC-G.114}},
  note         = {Accessed: 2025-04-10}
}

@inproceedings{ai2023neural,
  title={Neural speech phase prediction based on parallel estimation architecture and anti-wrapping losses},
  author={Ai, Yang and Ling, Zhen-Hua},
  booktitle={ICASSP 2023-2023 IEEE International Conference on Acoustics, Speech and Signal Processing (ICASSP)},
  pages={1--5},
  year={2023},
  organization={IEEE}
}

@article{ba2016layer,
  title={Layer normalization},
  author={Ba, Jimmy Lei and Kiros, Jamie Ryan and Hinton, Geoffrey E},
  journal={arXiv preprint arXiv:1607.06450},
  year={2016}
}

@article{kumar2019melgan,
  title={Melgan: Generative adversarial networks for conditional waveform synthesis},
  author={Kumar, Kundan and Kumar, Rithesh and De Boissiere, Thibault and Gestin, Lucas and Teoh, Wei Zhen and Sotelo, Jose and De Brebisson, Alexandre and Bengio, Yoshua and Courville, Aaron C},
  journal={Advances in neural information processing systems},
  volume={32},
  year={2019}
}

@article{kong2020hifi,
  title={Hifi-gan: Generative adversarial networks for efficient and high fidelity speech synthesis},
  author={Kong, Jungil and Kim, Jaehyeon and Bae, Jaekyoung},
  journal={Advances in neural information processing systems},
  volume={33},
  pages={17022--17033},
  year={2020}
}

\end{document}